\documentclass[notitlepage,aps,prl,reprint, groupedaddress]{revtex4-2}
\usepackage[utf8]{inputenc}

\usepackage{qcircuit}
\usepackage{mathtools}
\usepackage[english]{babel}
\usepackage[utf8]{inputenc}
\usepackage{amsmath}
\usepackage{amsfonts}
\usepackage{amssymb}
\usepackage{graphicx}
\usepackage{bbold}
\usepackage{epstopdf}
\usepackage{hyperref}
 
\renewcommand*{\Im}{\operatorname{Im}} 
\DeclarePairedDelimiter\bra{\langle}{\rvert}
\DeclarePairedDelimiter\ket{\lvert}{\rangle}

\begin{document}
\renewcommand{\figurename}{FIG.}
\renewcommand{\tablename}{TABLE}
\title{All-optical cat-code quantum error correction}
\author{Jacob Hastrup}
\email{jhast@fysik.dtu.dk}
\author{Ulrik Lund Andersen}
\email{ulrik.andersen@fysik.dtu.dk}

\affiliation{Center for Macroscopic Quantum States (bigQ), Department of Physics, Technical University of Denmark, Building 307, Fysikvej, 2800 Kgs. Lyngby, Denmark}

\begin{abstract}
The cat code is a promising encoding scheme for bosonic quantum error correction as it allows for correction against losses---the dominant error mechanism in most bosonic systems. However, for losses to be detected efficiently without disturbing the encoded logical information, one needs to implement a parity measurement of the excitation number. While such a measurement has been demonstrated in the microwave regime using a superconducting transmon ancilla, it has remained unclear how it can be implemented in the optical regime. Here, we introduce a teleportation-based error-correction scheme for the cat code, using elements suitable for an optical setting. The scheme detects and corrects single-photon losses while restoring the amplitude of the cat states, thereby greatly suppressing the accumulation of errors in lossy channels. 
\end{abstract}
\date{\today}

\maketitle

\section{Introduction}
Quantum states are notoriously vulnerable to external noise sources, posing a central challenge towards making useful quantum technologies. To overcome the effects of noise, numerous quantum error-correction protocols have been developed over the past 25 years. The main idea behind quantum error correction is to redundantly encode each logical qubit into a larger Hilbert space, such that noise can be detected before it accumulates into logical errors. The most common approach is to use multiple physical modes to encode each qubit. Alternatively, one can encode a qubit into multiple energy levels of a single bosonic mode \cite{Albert2018PerformanceCodes,Cochrane1999MacroscopicallyDamping, Ralph2003QuantumStates, Bergmann2016QuantumStates, Ofek2016ExtendingCircuits,Gottesman2001EncodingOscillator, Grimsmo2020QuantumCodes, Michael2016NewMode}. Such bosonic encoding can provide an advantage in terms of hardware efficiency, reducing the number of physical modes per logical qubit. Furthermore, bosonic error correction---and in particular the cat code---has been experimentally used to extend the lifetime of a qubit beyond what is achievable in the same system without error correction \cite{Ofek2016ExtendingCircuits}; a feat which remains to be demonstrated in conventional qubit systems. 

A prominent bosonic system is the electromagnetic field, which has seen much development towards quantum technologies in recent years in both the optical regime \cite{Takeda2019TowardComputing,Tzitrin2021Fault-tolerantOptics,Larsen2019DeterministicState,Asavanant2019GenerationState,Larsen2021DeterministicPlatform,Hacker2019DeterministicStates} as well as the microwave regime \cite{Campagne-Ibarcq2020QuantumOscillator,Reinhold2020Error-correctedQubit,Hu2019QuantumQubit,Ofek2016ExtendingCircuits}. In this work, we are concerned with the optical regime, which provides several advantages. For example, optical field modes are naturally in the vacuum state at room temperature, limiting the requirement for cryogenics. Additionally, optical modes are easily entangled using beamsplitters, enabling large-scale entanglement \cite{Yoshikawa2016InvitedMultiplexing,Larsen2019DeterministicState,Asavanant2019GenerationState}. Furthermore, travelling optical modes constitute an uncontested platform for quantum communication. 

However, optical losses can rapidly wash out vital quantum signatures. Therefore, the quantum states should be encoded such that small losses can be detected and corrected before they accumulate. The two most promising bosonic encodings against losses \cite{Albert2018PerformanceCodes} in the optical regime are the Gottesman-Kitaev-Preskill (GKP) code \cite{Gottesman2001EncodingOscillator,Tzitrin2020ProgressCodes,EliBourassa2021BlueprintComputer,Larsen2021Fault-TolerantArchitecture} and the cat code \cite{Bergmann2016QuantumStates, Li2017CatChannel, Leghtas2013Hardware-efficientProtection}. The GKP code, in particular, has in recent years gained renewed interest and numerous new developments for optical systems have been witnessed. This interest has largely been fueled by the potential ease of implementing gates and error correction, given a supply of high-quality encoded states. However, GKP-encoded states have yet to be produced in the optical regime, and theoretical analyses indicate that GKP states of useful quality will be challenging to produce using practical noisy components \cite{Tzitrin2020ProgressCodes,Hastrup2021GenerationQED}.

Meanwhile, optical two-component Schrödinger's cat states---which are the encoded states of the cat code---have already been produced experimentally \cite{Hacker2019DeterministicStates,Neergaard-Nielsen2006GenerationNetworks}. And while their superpositions, corresponding to four-components Schrödinger's cat states remain to be produced optically, several schemes have been proposed for this task \cite{Hastrup2020DeterministicState, Thekkadath2020EngineeringDetector, Asavanant2021Wave-functionResource}. But unlike the GKP codes, there has been no proposal on how to implement gates or perform error correction on cat codes using tools available in the optical regime, thereby limiting their use in practice. In this work, we address part of this issue by proposing an all-optical teleportation setup which allows for single-photon losses to be detected and corrected without disturbing the encoded logical information.  

\section{Protocol}

 \begin{figure}
    \centering
    \includegraphics{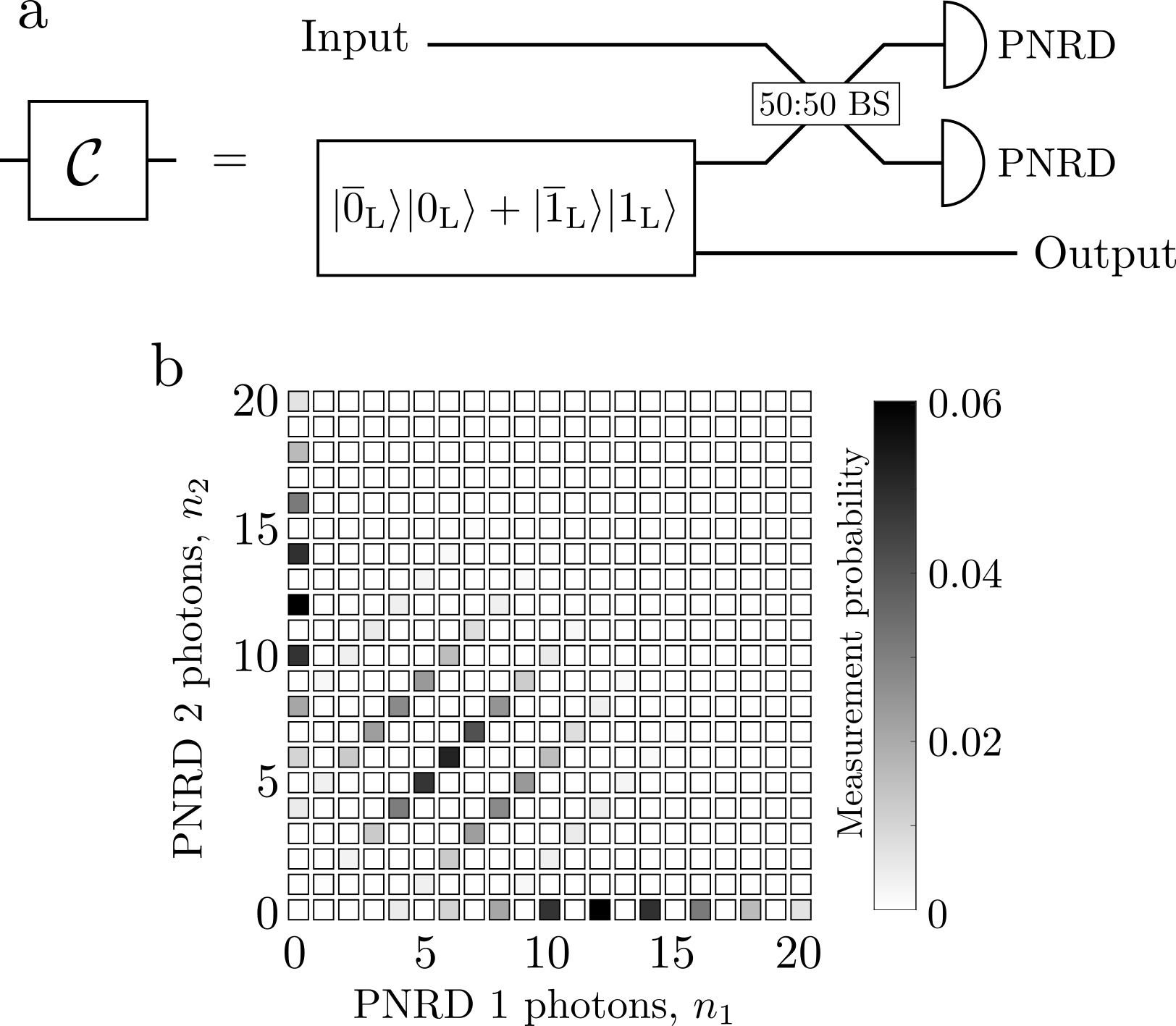}
    \caption{(a): Proposed error-correction circuit for performing all-optical error correction of the cat codes. The input state to be corrected is mixed on a 50:50 BS with one half of a logical Bell state and the output modes are subsequently measured with PNRDs. This teleports the input state to the other half of the Bell state while correcting single-photon losses. (b): Probability of obtaining measurement result $(n_1,n_2)$ for an input logical state with amplitude $\alpha = 2.5$, without losses. The results are either distributed with both modes around $\alpha^2$, or with one mode containing 0 photons and the other mode containing about $2\alpha^2$ photons, in accordance with Eq.\ \eqref{eq:postBS}. }
    \label{fig:setup}
\end{figure}

The logical basis states of the cat code are two-component Schrödinger's cat states given by:
\begin{align}
    \ket{0_\textrm{L}} &=  \frac{\ket{\alpha} + \ket{-\alpha}}{\sqrt{2(1 + e^{-2|\alpha|^2})}}, \\
    \ket{1_\textrm{L}} &=  \frac{\ket{i\alpha} + \ket{-i\alpha}}{\sqrt{2(1 + e^{-2|\alpha|^2})}},
\end{align}
where $\ket{\alpha}$ is a coherent state. The amplitude, $\alpha$, which we assume to be real, should be chosen to optimize the performance of the code, as we will discuss later. Note that due to the finite overlap between coherent states, $\langle \beta_1|\beta_2\rangle = e^{i\Im (\beta_1^*\beta_2)}e^{-|\beta_1 - \beta_2|^2/2}$, the logical basis states are generally not orthogonal, i.e. $\langle 0_\textrm{L} | 1_\textrm{L} \rangle = \cos(\alpha^2)/\cosh(\alpha^2)$. However, the exponential increase of the hyperbolic cosine causes the overlap to quickly vanish for $\alpha \gtrapprox 2$. 

The encoded states in the Fock basis are given by:
\begin{align}
    \ket{0_\textrm{L}} = \frac{1}{\cosh(\alpha^2)}\sum_{n=0}^{\infty}\frac{(\alpha^2)^{n}}{\sqrt{(2n)!}}\ket{2n},\\
    \ket{1_\textrm{L}} = \frac{1}{\cosh(\alpha^2)}\sum_{n=0}^{\infty}\frac{(-\alpha^2)^{n}}{\sqrt{(2n)!}}\ket{2n}.
\end{align}
Notably, the states have support only on every second photon-number state, which is a common property of states with a 2-fold phase-space symmetry \cite{Grimsmo2020QuantumCodes}. Therefore, we can detect if a single photon (or an odd number of photons) has been lost by measuring the photon-number parity of the state. Furthermore, the logical Pauli-X eigenstates $\ket{\pm_\textrm{L}} \propto \ket{0_\textrm{L}} \pm \ket{1_\textrm{L}}$ have support only on every fourth photon number, with $\ket{+_\textrm{L}}$ having support on $n \equiv 0$ (mod 4) and $\ket{-_\textrm{L}}$ having support on $n \equiv 2$ (mod 4). Thus a direct photon-number measurement realises a measurement in the logical X-basis. To detect photon loss without collapsing the logical state, we therefore need to extract information only on the parity without getting any information on the exact photon number.

This can be done using the error-correction circuit illustrated in Fig.\ \ref{fig:setup}a. It consists of a 50:50 beamsplitter (BS), two photon-number-resolving detectors (PNRDs), and an ancilla resource state in the form of a logical Bell state, $\ket{\overline{0}_\textrm{L}}\ket{0_\textrm{L}}+ \ket{\overline{1}_\textrm{L}}\ket{1_\textrm{L}}$, where the bar denotes cat states with a reduced amplitude that matches that of the input state after loss. In the Appendix we present a proposal on how this logical Bell state could be produced optically. The input state interferes with one half of the logical Bell state on the BS, the outputs of which are measured with the PNRDs. As a result of this measurement, the input state is teleported to the other half of the Bell state, while correcting for single-photon losses, i.e. Knill-type error correction is performed \cite{Knill2005ScalableRates}. Conceptually, since each mode of the Bell state contains an even number of photons the parity of the total number of photons measured by the PNRDs is determined by the parity of the input state, enabling the detection of losses. Meanwhile, since the state is mixed with the one half of the Bell state, we do not reveal information on the exact number of photons in the input state. Additionally, as the logical state is teleported onto a fresh cat-state ancilla, the cat-state amplitudes are restored to their initial values.

Before further discussions on the error-corrective properties of the circuit we interrogate the functionality of the qubit teleportation circuit in the absence of losses. Consider an arbitrary logical input state, $\mu\ket{0_\textrm{L}} + \nu\ket{1_\textrm{L}}$, written in the coherent state basis (neglecting normalization and with subscripts denoting the different modes):
\begin{widetext}
\begin{equation}
    \bigg[\mu(\ket{\alpha}_1 + \ket{-\alpha}_1) + \nu(\ket{i\alpha}_1 + \ket{-i\alpha}_1)\bigg]\bigg[(\ket{\alpha}_2 + \ket{-\alpha}_2)(\ket{\alpha}_3 + \ket{-\alpha}_3) + (\ket{i\alpha}_2 + \ket{-i\alpha}_2)(\ket{i\alpha}_3 + \ket{-i\alpha}_3)\bigg].
\end{equation}
As the BS transforms coherent states according to $\ket{\beta_1}_1\ket{\beta_2}_2 \rightarrow \ket{(\beta_1 + \beta_2)/\sqrt{2}}_1\ket{(-\beta_1 + \beta_2)/\sqrt{2}}_2$, the transformed state can be written in the form 
\begin{align}
    \rightarrow &\mu\bigg[\ket{\sqrt{2}\alpha}_1\ket{0}_2 + \ket{0}_1\ket{-\sqrt{2}\alpha}_2 + \ket{0}_1\ket{\sqrt{2}\alpha}_2 + \ket{-\sqrt{2}\alpha}_1\ket{0}_2\bigg]\ket{0_\textrm{L}}_3\nonumber \\
    &+ \nu\bigg[\ket{i\sqrt{2}\alpha}_1\ket{0}_2 + \ket{0}_1\ket{-i\sqrt{2}\alpha}_2 + \ket{0}_1\ket{i\sqrt{2}\alpha}_2 + \ket{-i\sqrt{2}\alpha}_1\ket{0}_2\bigg]\ket{1_\textrm{L}}_3 \nonumber \\
    &+ \mu\bigg[\ket{\tilde{\alpha}}_1\ket{-\tilde{\alpha}^*}_2 + \ket{\tilde{\alpha}^*}_1\ket{-\tilde{\alpha}}_2 + \ket{-\tilde{\alpha}^*}_1\ket{\tilde{\alpha}}_2 + \ket{-\tilde{\alpha}}_1\ket{\tilde{\alpha}^*}_2\bigg]\ket{1_\textrm{L}}_3\nonumber \\
    &+ \nu\bigg[\ket{\tilde{\alpha}}_1\ket{\tilde{\alpha}^*}_2 + \ket{-\tilde{\alpha}^*}_1\ket{-\tilde{\alpha}}_2 + \ket{\tilde{\alpha}^*}_1\ket{\tilde{\alpha}}_2 + \ket{-\tilde{\alpha}}_1\ket{-\tilde{\alpha}^*}_2\bigg]\ket{0_\textrm{L}}_3, \label{eq:postBS}
\end{align}
\end{widetext}
where $\tilde{\alpha} = (\alpha + i\alpha)/\sqrt{2}$ and $^*$ denotes complex conjugation. Two distinct cases appear: Either, one of the modes 1 and 2 is in the vacuum state with the other in a coherent state of magnitude $\sqrt{2}\alpha$, or both modes are in coherent states of magnitude $|\tilde{\alpha}|=\alpha$. In the regime where $\alpha$ is not too small, these two cases are distinguished by the PNRD results, as illustrated in Fig.\ \ref{fig:setup}b. In the first case, the coefficients $\mu$ and $\nu$ get mapped to the correct corresponding logical state in mode 3. In the second case, the coefficients get swapped, in which case a corrective logical $X$ gate should be applied. Further analysis (see Appendix) shows that measurement results with a total photon number of $2$ (mod 4) add a -1 phase to the $\nu$ term, thus requiring a logical $Z$ gate correction. In summary, the circuit performs a teleportation of mode 1 onto mode 3, with the measurement result signaling which Pauli correction should be applied to mode 3 to retrieve the input state, just like standard qubit teleportation. 

An important caveat should be noted: Knowing whether a logical $X$ correction should be applied relies on the fact that we can distinguish the first two lines of Eq.\ \eqref{eq:postBS} from the last two with the PNRDs. If $\alpha$ is too small, the probability of measuring e.g.\ $(n_1,n_2) = (0,4)$ has contributions from both cases. That is, the central contributions in Fig.\ \ref{fig:setup}b overlap with the edge cases. It turns out that the total probability in this case depends on the input state, and in particular, such measurement results become a weak unwanted logical $X$ measurement (See Appendix for details). To intuitively understand this, consider the extreme case of measuring $(n_1,n_2) = (0,0)$. This can only occur if both arms before the BS have support on the vacuum state. But the state $\ket{-_L}$ does not have support on the vacuum, and so this measurement projects the input state onto $\ket{+_L}$. This can thus effectively cause logical depolarization of e.g.\ logical $Z$ or $Y$ states. To avoid this, we should choose large enough $\alpha$, such that this measurement result becomes unlikely. 

To characterize the performance of the circuit, we consider how well input logical Pauli-eigenstates remain distinguishable. That is, we consider three sets of input states, $\{\ket{0_\textrm{L}}, \ket{1_\textrm{L}} \}$, $\{\ket{+_\textrm{L}}, \ket{-_\textrm{L}} \}$ and $\{\ket{+i_\textrm{L}}, \ket{-i_\textrm{L}} \}$, where $\ket{\pm i_\textrm{L}}\propto \ket{0_\textrm{L}} \pm i\ket{0_\textrm{L}}$ denotes the logical Pauli-Y eigenstates. Denoting the channel realized by the error-correction circuit by $\mathcal{C}$, averaging over the PNRD measurement results weighted according to their probability distribution and keeping track of any induced Pauli-rotations, we calculate the resulting, generally mixed, output states, e.g.\ $\rho^{Z}_0 = \mathcal{C}(\ket{0_\textrm{L}})$ and $\rho^{Z}_1 = \mathcal{C}(\ket{1_\textrm{L}})$. We then calculate the probability of misidentifying the input $0$ state as a $1$ or vice versa, which is bounded by the Helstrom bound \cite{Helstrom1976QuantumTheory}:
\begin{equation}
    p_\textrm{err}^Z = \frac{1}{2} - \frac{1}{4}(||\rho_0^Z - \rho_1^Z||_1),
\end{equation}
where $||\cdot||_1$ denotes the trace norm. We consider the Helstrom bound to focus our attention to the intrinsic properties of the circuit. 

Similarly, we denote the error probabilities of the Pauli $X$ and $Y$ states as $p_\textrm{err}^X$ and $p_\textrm{err}^Y$. We then define the average of these error rates as our figure of merit,
\begin{equation}
    p_\textrm{err} \equiv (p_\textrm{err}^X + p_\textrm{err}^Y + p_\textrm{err}^Z)/3.
\end{equation} 
A value of $p_\textrm{err} = 0.5$ thus corresponds to a complete loss of the logical information, while $p_\textrm{err} = 0$ is a perfect preservation of the information.

\begin{figure}
    \centering
    \includegraphics{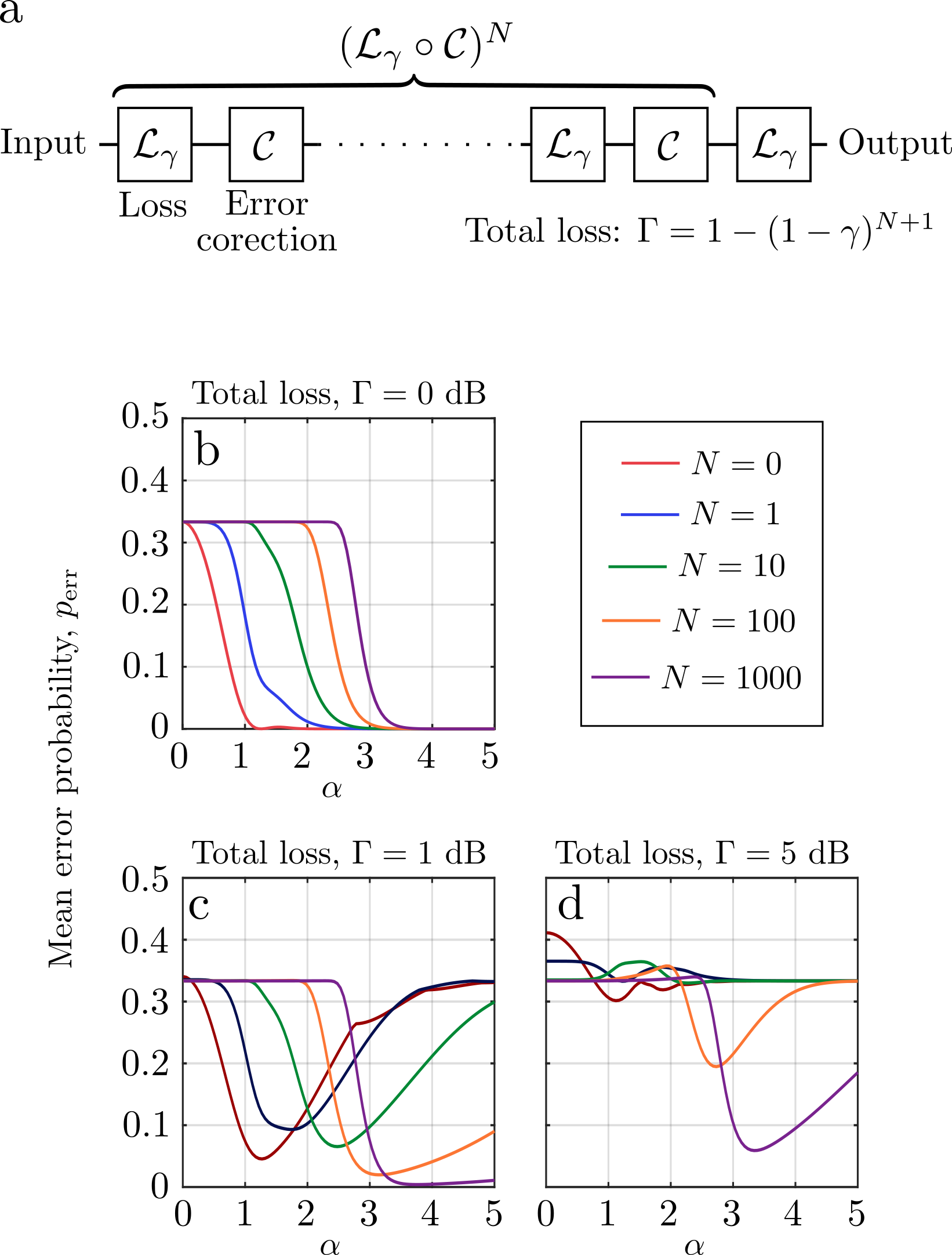}
    \caption{(a): The lossy channel $\mathcal{L}_\Gamma$ is divided into smaller segments, $\mathcal{L}_\gamma$ with $N$ error correction steps. (b): Minimum measurement error probability of input logical Pauli states after $N$ applications of the correction circuit of Fig.\ \ref{fig:setup}a, averaging over all possible PNRD measurement outcomes. $N=0$ corresponds to the identity channel. (c,d): Minimum error probability after a lossy channel divided into $N+1$ smaller segments with $N$ correction circuits distributed between the lossy segments.}
    \label{fig:alpha}
\end{figure}

Fig.\ \ref{fig:alpha}b shows the performance of the error-correction circuit in the absence of any losses, repeatedly applying the circuit $N$ times to the same state. The red $N = 0$ curve thus simply corresponds to the identity channel. For this curve, we note that $p_\textrm{err}$ goes to 1/3 for small $\alpha$. This is due to the indistinguishability of the logical $Z$ and $Y$ states, which all converge to the vacuum state in the limit of small $\alpha$. The logical $X$ states, however, remain perfectly distinguishable for all non-zero $\alpha$, resulting in a total average error rate of $2/3 \times 0.5 = 1/3$. 

When applying the teleportation circuit, $p_\textrm{err}$ increases at small $\alpha$ due to the weak Pauli $X$ measurement discussed earlier. This effect accumulates when applying the circuit multiple times. However, if $\alpha$ is sufficiently large, e.g.\ at $\alpha=4$, this effect becomes negligible even when the circuit is applied many times. 

\begin{figure}
    \centering
    \includegraphics{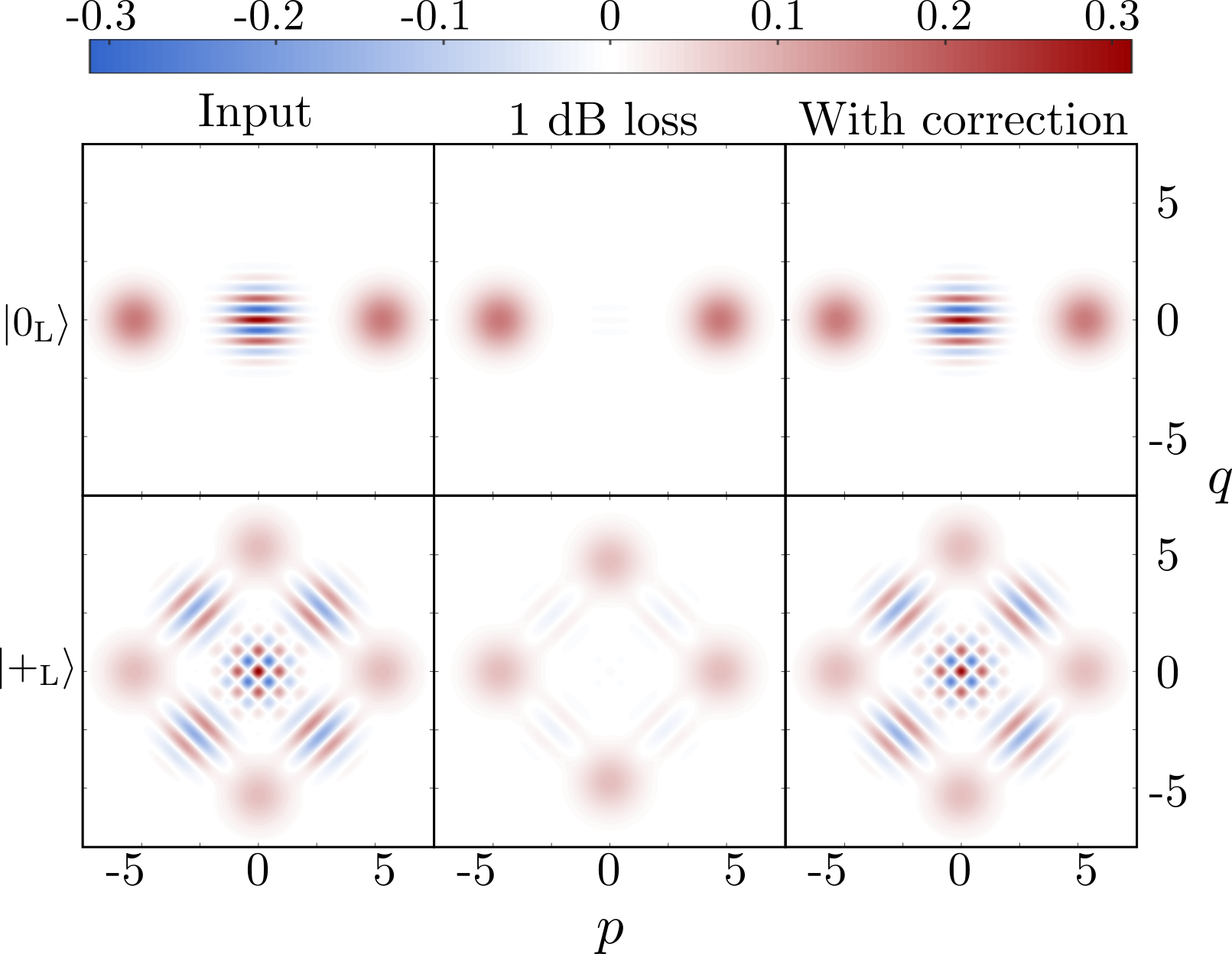}
    \caption{Wigner functions for cat states with $\alpha = 3$ before loss (input), after $\Gamma=1$ dB loss without error correction, and $\Gamma = 1$ dB loss divided by $N=100$ error correction steps. Top row: input logical $Z$ state. $\ket{0_L}$. Bottom row: input logical $X$ state, $\ket{+_L}$. The shown error-corrected states are averaged over PNRD measurement results, and post-selected such that the required corrective Pauli rotation adds up to the identity.} 
    \label{fig:wigner}
\end{figure}

We now consider the case of lossy input states. Losses of magnitude $\Gamma$ are described by the channel
\begin{equation}
    \mathcal{L}_\Gamma(\rho) = \sum_{l=0}^\infty \hat{K}^{(l)}_\Gamma \rho (\hat{K}^{(l)}_\Gamma)^\dagger,
\end{equation}
with Kraus operators:
\begin{equation}
    \hat{K}^{(l)}_\Gamma = \sqrt{\frac{\Gamma}{1-\Gamma}}^{l}\frac{\hat{a}^l}{\sqrt{l!}}\sqrt{1-\Gamma}^{\hat{n}}.
\end{equation}
For a fixed total loss, $\Gamma$, we break the channel into $N+1$ equal segments using $N$ correction circuits, yielding the total channel $(\mathcal{L}_\gamma \circ \mathcal{C})^N\circ \mathcal{L}_\gamma$ with segment loss $\gamma = 1 - \sqrt[N+1]{1 - \Gamma}$, as shown if Fig.\ \ref{fig:alpha}a. 

Fig.\ \ref{fig:alpha}c and d show the performance for $\Gamma = 1$ dB and $\Gamma = 5$ dB respectively, with units in dB defined as $-10\log_{10}(1 - \Gamma)$. For non-zero losses we now observe an optimum value of $\alpha$, depending on $N$. This is because states with large $\alpha$ contain more photons on average, and are thus more sensitive to losses. The error-correction circuit can correct single-photon losses, but not two-photon losses, and so $\alpha$ should be small enough to make two-photon losses improbable. Importantly, we see that repeated application of the correction circuit throughout the channel can significantly suppress the effects of loss. 

As seen in Fig.\ \ref{fig:alpha}c and d, more frequent error correction, i.e. larger $N$, allows for fewer errors by suitably optimizing $\alpha$. However, in practise we are limited in how frequently we can perform error correction. This limitation can be due to finite amounts of hardware but also due to finite losses introduced by the imperfect components of the error correction circuit itself. 

Fig.\ \ref{fig:wigner} compares the Wigner functions of the input states $\ket{0_L}$ and $\ket{+_L}$, with and without error correction for $\Gamma = 1$ dB total loss and $N=100$ error-correction steps. We see that the states, and in particular the negativities, are well-preserved by the error-correction protocol.

\begin{figure}
    \centering
    \includegraphics{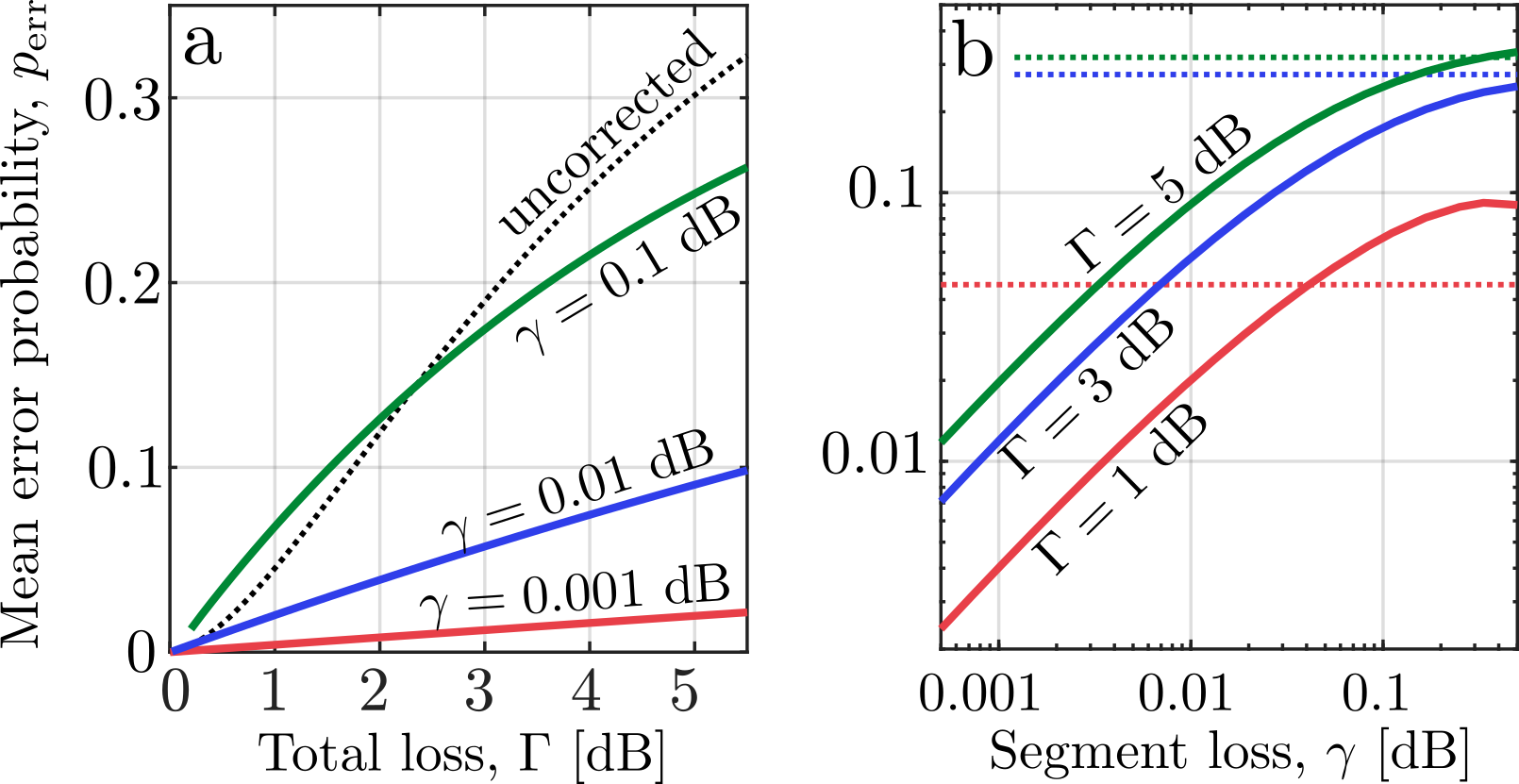}
    \caption{(a): Minimum error probability as a function of total channel loss, when performing error correction after every 0.1 dB (green), 0.01 dB (blue) or 0.001 dB (red) loss. The dotted line corresponds to direct transmission of a logical cat state through the channel. For all curves, $\alpha$ is chosen to minimize $p_\textrm{err}$. (b): Minimum error probability as a function of the loss between error-correction steps for a total amount of loss of $5$ dB (green), $3$ dB (blue) and $1$ dB (red). The dotted lines show the corresponding uncorrected cases.}
    \label{fig:gamma}
\end{figure}

Fig.\ \ref{fig:gamma}a shows how well the error-correction circuit protects against loss when optimizing over $\alpha$, with an amount of loss between error-correction steps of $0.1$ dB (2.3\%), $0.01$ dB (0.23\%) and $0.001$ dB (0.023\%). Comparing to the uncorrected case (dotted line), we see that error correction needs to be applied quite frequent to gain an advantage. 

In addition to loss correction in quantum communication, the circuit can also be relevant for loss-correction in optical quantum computing. In particular, we can imagine a temporal measurement-based computation scheme similar to what has been demonstrated in continuous-variable optics \cite{Yoshikawa2016InvitedMultiplexing,Larsen2021DeterministicPlatform}, where a single set of detectors can be used to perform arbitrarily many subsequent teleportations to repeatedly error-correct an encoded bosonic qubit. Of course, for such a scheme to be useful, we also need to be able to implement gates on the encoded state, which is outside the scope of this work. 

The more frequently we apply error correction, the faster we might accumulate errors due to the intrinsic weak Pauli $X$ measurement of the circuit, as was shown in Fig.\ \ref{fig:alpha}b. To suppress this effect we require larger $\alpha$, which in turn results in more losses. It is therefore relevant to ask whether this trade-off is favourable, i.e. how low can the total error rate be. This is examined in Fig.\ \ref{fig:gamma}b, showing $p_\textrm{err}$ as a function of the segment loss, $\gamma$, for different total loss channels, optimizing $\alpha$ for all curves. On the log-log plot we observe a linear relationship between $p_\textrm{err}$ and $\gamma$, indicating that the error can indeed in principle be made arbitrarily low given frequent enough error correction and a suitable choice of $\alpha$. In practise this relationship will be limited by hardware constraints, such as detector inefficiencies and Bell state preparation inefficiencies, which need to be negligible compared to $\gamma$ for Fig.\ \ref{fig:gamma} to hold. 

\section{Conclusion}
We have presented an all-optical scheme for performing quantum error correction on bosonic cat-code qubits, allowing single-photon loss events to be detected and corrected. The scheme relies on photon counting and logical Bell state resources and works like a conventional teleportation scheme in the absence of loss. For small cat-state amplitudes, some measurement results of the protocol act as a weak unwanted logical Pauli $X$ measurement, thereby effectively inducing depolarization around the $X$-axis. To minimize this effect, $\alpha$ should be optimized accordingly.

While frequent error correction is needed to effectively suppress errors using the proposed scheme, the errors can in principle be made arbitrarily low, greatly surpassing the uncorrected state. This work thus constitutes a step towards optical cat-based fault-tolerant quantum computations. Important future work will be to find methods of optically implementing gates on the encoded states, and improving the quality of experimentally generated cat-states. 

\section{Acknowledgements}
This  project  was  supported  by  the  Danish  National Research  Foundation  through  the  Center  of  Excellence for Macroscopic Quantum States (bigQ, DNRF0142).

\begin{widetext}
\section{Appendix A: Teleportation without losses}
As stated in the main text, the state after BS interaction is given by:
\begin{align}
    &\hat{U}_{BS}\left(\mu\ket{0_\textrm{L}}_1 + \nu\ket{1_\textrm{L}}_1\right)(\ket{0_\textrm{L}}_2\ket{0_\textrm{L}}_3 + \ket{1_\textrm{L}}_2\ket{1_\textrm{L}}_3) \nonumber\\
    =\, &\mu\bigg[\ket{\sqrt{2}\alpha}_1\ket{0}_2 + \ket{0}_1\ket{-\sqrt{2}\alpha}_2 + \ket{0}_1\ket{\sqrt{2}\alpha}_2 + \ket{-\sqrt{2}\alpha}_1\ket{0}_2\bigg]\ket{0_\textrm{L}}_3\nonumber \\
    &+ \nu\bigg[\ket{i\sqrt{2}\alpha}_1\ket{0}_2 + \ket{0}_1\ket{-i\sqrt{2}\alpha}_2 + \ket{0}_1\ket{i\sqrt{2}\alpha}_2 + \ket{-i\sqrt{2}\alpha}_1\ket{0}_2\bigg]\ket{1_\textrm{L}}_3 \nonumber \\
    &+ \mu\bigg[\ket{\tilde{\alpha}}_1\ket{-\tilde{\alpha}^*}_2 + \ket{\tilde{\alpha}^*}_1\ket{-\tilde{\alpha}}_2 + \ket{-\tilde{\alpha}^*}_1\ket{\tilde{\alpha}}_2 + \ket{-\tilde{\alpha}}_1\ket{\tilde{\alpha}^*}_2\bigg]\ket{1_\textrm{L}}_3\nonumber \\
    &+ \nu\bigg[\ket{\tilde{\alpha}}_1\ket{\tilde{\alpha}^*}_2 + \ket{-\tilde{\alpha}^*}_1\ket{-\tilde{\alpha}}_2 + \ket{\tilde{\alpha}^*}_1\ket{\tilde{\alpha}}_2 + \ket{-\tilde{\alpha}}_1\ket{-\tilde{\alpha}^*}_2\bigg]\ket{0_\textrm{L}}_3,
\end{align}
We now consider how mode 3 is projected when we measure $n_1$ and $n_2$ photons in modes 1 and 2 respectively. For this we make use of the relation
\begin{equation}
    \langle n|\beta\rangle=e^{-|\beta|^2/2}\frac{\beta^n}{\sqrt{n!}}.
\end{equation}
First, we consider the case when $n_1\neq 0$ and $n_2\neq 0$, in which case only the last two lines contribute. Recalling that $\tilde{\alpha}^* = -i\tilde{\alpha}$ we get:
\begin{align}
    &\bra{n_1}_1\bra{n_2}_2\hat{U}_{BS}\left(\mu\ket{0_\textrm{L}}_1 + \nu\ket{1_\textrm{L}}_1\right)(\ket{0_\textrm{L}}_2\ket{0_\textrm{L}}_3 + \ket{1_\textrm{L}}_2\ket{1_\textrm{L}}_3) \nonumber \\
    = & \, \frac{e^{-|\alpha|^2}}{\sqrt{n_1!n_2!}}\tilde{\alpha}^{n_1 + n_2}(i^{n_1}+i^{n_2})(1 + (-1)^{n_1 + n_2})\left(\mu\ket{1_\textrm{L}}_3 + \frac{(-1)^{n_1}+(-1)^{n_2}}{2}\nu\ket{0_\textrm{L}}_3\right).
\end{align}
There are a few things to note in this expression: First, the prefactor $1 + (-1)^{n_1 + n_2}$ yields $0$ whenever $n_1$ and $n_2$ are of different parity, i.e. when the total photon number is odd. Thus we are guaranteed to measure an even total photon number, as expected since the input states contains only even photon numbers. Second, the prefactor $i^{n_1} + i^{n_2}$ yields 0 whenever $n_1$ and $n_2$ are different modulus 4, assuming equal parity of $n_1$ and $n_2$. This is less trivial, and a result of quantum interference on the BS. This can also be seen in Fig.\ \ref{fig:setup}b. Third, the coefficients get swapped, i.e. $\mu$ is mapped onto $\ket{1_\textrm{L}}$ and $\nu$ to $\ket{0_\textrm{L}}$, requiring a corrective logical $X$ gate, as stated in the main text. Finally, the $[(-1)^{n_1} + (-1)^{n_2}]/2$ term contributes a $(-1)$ phase factor whenever $n_1$ and $n_2$ are odd, requiring a corrective logical $Z$ gate. Since $n_1\equiv n_2$ (mod 4) this corresponds to the cases when $n_1 + n_2 \equiv 2$ (mod 4). 

Next, we consider the case when $n_2 = 0$ and $n_1 \neq 0$:
\begin{align}
    &\bra{n_1}_1\bra{0}_2\hat{U}_{BS}\left(\mu\ket{0_\textrm{L}}_1 + \nu\ket{1_\textrm{L}}_1\right)(\ket{0_\textrm{L}}_2\ket{0_\textrm{L}}_3 + \ket{1_\textrm{L}}_2\ket{1_\textrm{L}}_3) \nonumber \\
    = & \, \frac{e^{-|\alpha|^2}}{\sqrt{n_1!}}(\sqrt{2}\alpha)^{n_1}(1 + (-1)^{n_1})\left(\mu\ket{0_\textrm{L}}_3 + i^{n_1}\nu\ket{1_\textrm{L}}_3\right) \nonumber \\
    & + \frac{e^{-|\alpha|^2}}{\sqrt{n_1!}}\tilde{\alpha}^{n_1}(i^{n_1}+1)(1 + (-1)^{n_1})\left(\mu\ket{1_\textrm{L}}_3 + \frac{(-1)^{n_1}+1}{2}\nu\ket{0_\textrm{L}}_3\right) \nonumber \\
    = & \, \frac{e^{-|\alpha|^2}}{\sqrt{n_1!}}\sqrt{2}^{n_1}\alpha^{n_1}(1 + (-1)^{n_1})\left[ \left(\mu + \frac{e^{in_1\pi/4}}{\sqrt{2}^{n_1}}(i^{n_1}+1)\frac{(-1)^{n_1} + 1}{2}\nu \right)\ket{0_\textrm{L}}_3 + \left(i^{n_1}\nu + (i^{n_1}+1)\frac{e^{in_1\pi/4}}{\sqrt{2}^{n_1}}\mu\right)\ket{1_\textrm{L}}_3\right].
\intertext{When $n_1 \equiv 2$ (mod 4) this reduces to:}
    \propto & \left[\mu \ket{0_\textrm{L}}_3 -\nu\ket{1_\textrm{L}}_3\right].
\intertext{That is, we recover the state with a corrective logical $Z$ gate. When $n_2 \equiv 0$ (mod 4) we get:}
    \propto & \left[\left(\mu + \frac{(-1)^{n_1/4}}{\sqrt{2}^{n_1-2}}\nu\right) \ket{0_\textrm{L}}_3 + \left(\nu + \frac{(-1)^{n_1/4}}{\sqrt{2}^{n_1-2}}\mu\right)\ket{1_\textrm{L}}_3\right].
\end{align}
In this case we induce a non-correctable error on the output state. However, when $n_1$ is large, the expression reduces to $\mu\ket{0_\textrm{L}}_3 + \nu\ket{1_\textrm{L}}_3$, i.e.\ we recover the state without error. On the other hand, when $n_1=4$, for example, we get $\left(\mu -0.5\nu\right)\ket{0_\textrm{L}}_3 + \left(\nu - 0.5\mu\right)\ket{1_\textrm{L}}_3$. In this case, the amplitudes are reduced when $\mu$ and $\nu$ are of the same sign, and increased when their signs differ. Thus this measurement outcome is more likely for $\ket{-_\textrm{L}}$ states compared to $\ket{+_\textrm{L}}$ states, i.e. the measurement becomes a weak logical Pauli $X$ measurement. To avoid this, $\alpha$ should be chosen sufficiently large to reduce the probability of this measurement outcome for all logical input states.

Finally, we consider the case when $n_1 = n_2 = 0$:
\begin{align}
    &\bra{0}_1\bra{0}_2\hat{U}_{BS}\left(\mu\ket{0_\textrm{L}}_1 + \nu\ket{1_\textrm{L}}_1\right)(\ket{0_\textrm{L}}_2\ket{0_\textrm{L}}_3 + \ket{1_\textrm{L}}_2\ket{1_\textrm{L}}_3) \nonumber \\
    = &\, 4e^{-|\alpha|^2}(\mu + \nu)\left[\ket{0_\textrm{L}}_3 + \ket{1_\textrm{L}}_3  \right].
\end{align}
In this case, the output is completely independent of the input. Furthermore, this outcome does not occur when $\mu = -\nu$, and thus the result projects the output onto the $\ket{+_\textrm{L}}$ state. Again, to avoid this scenario $\alpha$ should be chosen sufficiently large.

\section{Appendix B: Teleportation with losses}
We now look more detailed at the loss channel. As described in the main text, loss is described by Kraus operators
\begin{equation}
    \hat{K}^{(l)}_\Gamma = \sqrt{\frac{\Gamma}{1-\Gamma}}^{l}\frac{\hat{a}^l}{\sqrt{l!}}\sqrt{1-\Gamma}^{\hat{n}},
\end{equation}
where the $l$'th Kraus operator $K^{(l)}_\Gamma$ corresponds to the case of loosing $l$ photons. The effect of this operator on the encoded state is:
\begin{align}
    \hat{K}_\Gamma^{(l)}\left(\mu\ket{0_\textrm{L}} + \nu\ket{1_\textrm{L}}\right) &=  \sqrt{\frac{\Gamma}{1-\Gamma}}^{l}\frac{\hat{a}^l}{\sqrt{l!}} \left[\mu \left(\ket{\alpha\sqrt{1 - \Gamma}}+\ket{-\alpha\sqrt{1 - \Gamma}}\right) + \mu \left(\ket{i\alpha\sqrt{1 - \Gamma}}+\ket{-i\alpha\sqrt{1 - \Gamma}}\right)\right] \nonumber \\
    & \propto \hat{a}^l\left(\mu\ket{\overline{0}_L} + \nu\ket{\overline{1}_L}\right),
\end{align}
where $\ket{\overline{0}_L}$ and $\ket{\overline{1}_L}$ denote logical states with the reduced amplitude $\alpha_\Gamma \equiv \alpha\sqrt{1 - \Gamma}$. The $l=0$ term is corrected by the circuit as described in Appendix A, by scaling the amplitude of the first mode of the Bell state accordingly to $\alpha_\Gamma$. 

For $l=1$ we get the state:
\begin{equation}
\hat{K}_\Gamma^{(1)}\left(\mu\ket{0_\textrm{L}} + \nu\ket{1_\textrm{L}}\right) \propto \mu\left(\ket{\alpha_\Gamma} - \ket{-\alpha_\Gamma}\right) + i\nu \left(\ket{i\alpha_\Gamma} - \ket{-i\alpha_\Gamma}\right).
\end{equation}
Repeating the calculations of Appendix A, keeping track of the new (-1) signs, we find the output state for the different photon-number measurement results. When $n_1\neq 0 $ and $n_2\neq 0 $ we get:
\begin{align}
    &\bra{n_1}_1\bra{n_2}_2\hat{U}_{BS}\hat{a}\left(\mu\ket{\overline{0}_\textrm{L}}_1 + \nu\ket{\overline{1}_\textrm{L}}_1\right)(\ket{\overline{0}_\textrm{L}}_2\ket{0_\textrm{L}}_3 + \ket{\overline{1}_\textrm{L}}_2\ket{1_\textrm{L}}_3) \nonumber \\
    = & \, \frac{\alpha_\Gamma e^{-|\alpha_\Gamma|^2}}{\sqrt{n_1!n_2!}}\tilde{\alpha}^{n_1 + n_2}_\Gamma(i^{n_2}-i^{n_1})(1 - (-1)^{n_1 + n_2})\left(\mu\ket{1_\textrm{L}}_3 + \frac{i}{2}\left((-1)^{n_1} + (-1)^{n_2} + 2i^{n_1 + n_2}\right)\nu\ket{0_\textrm{L}}_3\right).
\end{align}
This time we find a contribution only when $n_1+n_2$ is odd, as expected. Again, a corrective $X$ gate should be applied when $n_1 \neq 0$ and $n_2 \neq 0$. Additionally, as we only need to consider odd $n_1+n_2$, the factor on the $\ket{0_\textrm{L}}_3$ term reduces to $(-1)^{(n_1+n_2+1)/2}$, i.e., a corrective $Z$ rotation is required when $n_1 + n_2\equiv 1$ (mod 4). 

When $n_1 \neq 0$ and $n_2 = 0$:
\begin{align}
    &\bra{n_1}_1\bra{0}_2\hat{U}_{BS}\hat{a}\left(\mu\ket{\overline{0}_\textrm{L}}_1 + \nu\ket{\overline{1}_\textrm{L}}_1\right)(\ket{\overline{0}_\textrm{L}}_2\ket{0_\textrm{L}}_3 + \ket{\overline{1}_\textrm{L}}_2\ket{1_\textrm{L}}_3) \nonumber \\
    = & \, \frac{\alpha_\Gamma e^{-|\alpha_\Gamma|^2}}{\sqrt{n_1!}}(\sqrt{2}\alpha_\Gamma)^{n_1}(1 - (-1)^{n_1})\left(\mu\ket{0_\textrm{L}}_3 + i^{n_1+1}\nu\ket{1_\textrm{L}}_3\right) \nonumber \\
    & + \frac{\alpha_\Gamma e^{-|\alpha_\Gamma|^2}}{\sqrt{n_1!}}\tilde{\alpha}_\Gamma^{n_1}(1 - i^{n_1})(1 - (-1)^{n_1})\left(\mu\ket{1_\textrm{L}}_3 + (-1)^{(n_1+1)/2}\nu\ket{0_\textrm{L}}_3\right) \nonumber  \\
    = & \, \frac{\alpha_\Gamma e^{-|\alpha_\Gamma|^2}}{\sqrt{n_1!}}(\sqrt{2}\alpha_\Gamma)^{n_1}(1 - (-1)^{n_1})\left[\left(\mu + \frac{e^{in_1\pi/4}}{\sqrt{2}^{n_1}}(1-i^{n_1})(-1)^{(n_1+1)/2}\nu   \right)\ket{0_\textrm{L}}_3 + \left(i^{n_1 + 1}\nu + \frac{e^{in_1\pi/4}}{\sqrt{2}^{n_1}}(1 - i^{n_1})\mu\right)\ket{1_\textrm{L}}_3  \right].
    \intertext{When $n_1 \equiv 1$ (mod 4):}
    \propto & \left[\left(\mu - \frac{e^{i\pi/4(n_1 -1)}}{\sqrt{2}^{n_1 + 1}} \nu\right)\ket{0_\textrm{L}}_3 + \left(-\nu + \frac{e^{i\pi/4(n_1 - 1)}}{\sqrt{2}^{n_1 + 1}}\mu\right)\ket{1_\textrm{L}}_3\right].
    \intertext{When $n_1 \equiv 3$ (mod 4):}
    \propto & \left[\left(\mu + \frac{e^{i\pi/4(n_1 +1)}}{\sqrt{2}^{n_1 + 1}} \nu\right)\ket{0_\textrm{L}}_3 + \left(\nu + \frac{e^{i\pi/4(n_1 + 1)}}{\sqrt{2}^{n_1 + 1}}\mu\right)\ket{1_\textrm{L}}_3\right].
\end{align}
Thus we get an uncorrectable contribution in both cases. However, for large $n_1$ this error vanishes due to the factor $\sqrt{2}^{n_1 + 1}$, and the input state is recovered by applying a logical $Z$ correction when $n_1 \equiv 1$ (mod 4). 
Finally, the case $n_1 = n_2 = 0$ occurs with probability $0$, as we expect to measure an odd number of photons if we have lost one photon in the input state. 

For $l=2$ we get 
\begin{equation}
\hat{K}_\Gamma^{(2)}\left(\mu\ket{0_\textrm{L}} + \nu\ket{1_\textrm{L}}\right) \propto \mu\ket{\overline{0}_\textrm{L}}- \nu \ket{\overline{1}_\textrm{L}}.
\end{equation}
That is, two-photon loss corresponds to an undetectable logical $Z$ rotation. 

To summarise, we can distinguish between the cases $l=0$ and $l=1$ by the parity of the total number of photons detected, $n_1 + n_2$. In both cases, a logical $X$ correction should be applied whenever both detectors measure more than $0$ photons. Additionally, a logical $Z$ correction should be applied whenever the total photon number modulus 4 is 1 or 2. 

\section{Appendix C: generation of logical Bell states with cavity QED}
Here we propose a method to generate the required logical Bell states. This is inspired by the experiment by Hacker et al. \cite{Hacker2019DeterministicStates}, which used a cavity QED system to generate a two-component cat state, $\ket{\alpha} + \ket{-\alpha}$. The cavity QED system consists of a cavity containing an atom with a three-level energy structure, $\ket{\downarrow},\ket{\uparrow}$ and $\ket{e}$, where the $\ket{\uparrow} \leftrightarrow \ket{e}$ transition is resonant with the cavity. By preparing the atom in the $(\ket{\downarrow},\ket{\uparrow})$-subspace and reflecting a coherent state pulse off the cavity, the reflected state obtains a phase shift depending on the state of the atom. In particular, the reflection coefficients can be written as (see Supplementary Information of Ref.\ \cite{Hacker2019DeterministicStates})
\begin{equation}
    r_\uparrow = 1 - \frac{2\kappa_r(i\Delta + \gamma)}{(i\Delta + \kappa)(i\Delta + \gamma) + g^2}, \qquad r_\downarrow = 1 - \frac{2\kappa_r}{i\Delta + \kappa}, \label{eq:reflection}
\end{equation}
where $\kappa_r$ is the coupling rate between the input free-space mode and the cavity, $\kappa$ is the total decay rate of the cavity, $\Delta$ is the detuning between the input field and the cavity, $\gamma$ is the spontaneous decay rate of the state $\ket{e}$ via modes other than the cavity mode and $g$ is the coupling strength between the atom and the cavity. 

When $\Delta=0$ and in the regime where $\kappa\approx \kappa_r$ and $g^2\gg \kappa\gamma$ we get $r_\uparrow = 1$ and $r_\downarrow = -1$, so for the atom in the initial state $\ket{+} = (\ket{\uparrow} + \ket{\downarrow})/\sqrt{2}$ a reflected coherent state becomes $\ket{\alpha} \rightarrow (\ket{\alpha}\ket{\uparrow} + \ket{-\alpha}\ket{\downarrow})\sqrt{2}$. Subsequently measuring the atom in the state $\ket{+}$ projects the optical field into the cat state $\propto \ket{\alpha} + \ket{-\alpha}$. Additionally, according to Eq.\ \eqref{eq:reflection} and as was experimentally demonstrated in \cite{Hacker2019DeterministicStates}, for $\Delta \neq 0$ we can obtain various phases between the coefficients $r_\uparrow$ and $r_\downarrow$. In particular, for $\Delta \approx \kappa$ (still in the regime $\kappa\approx\kappa_r$ and $g^2\gg \kappa \gamma$ and further assuming $g\gg \kappa$) we get a $\pi/2$ phase shift between the reflection coefficients. Thus, if a cat state is reflected onto this detuned cavity with the atom prepared in the $\ket{+}$ state, we obtain the state 
\begin{equation}
    \ket{\alpha} + \ket{-\alpha} \rightarrow \frac{1}{\sqrt{2}}(\ket{\alpha} + \ket{-\alpha})\ket{\uparrow} + \frac{1}{\sqrt{2}}(\ket{i\alpha} + \ket{-i\alpha})\ket{\downarrow},
\end{equation}
where the output state is written in a frame such that $\alpha$ is real. Measuring the atom now projects the optical field onto a 4-component cat state. However, if we instead reflect a second cat state off the cavity we get:
\begin{align}
    &\frac{1}{\sqrt{2}}\Big[(\ket{\alpha}_1 + \ket{-\alpha}_1)\ket{\uparrow} + (\ket{i\alpha}_1 + \ket{-i\alpha}_1)\ket{\downarrow}\Big]\Big[\ket{\alpha}_2 + \ket{-\alpha}_2\Big] \nonumber\\
    &\rightarrow \frac{1}{\sqrt{2}}\Big[(\ket{\alpha}_1 + \ket{-\alpha}_1)(\ket{\alpha}_2 + \ket{-\alpha}_2)\ket{\uparrow} + (\ket{i\alpha}_1 + \ket{-i\alpha}_1)(\ket{i\alpha}_2 + \ket{-i\alpha}_2)\ket{\downarrow}\Big].
\intertext{Measuring the atom in the $\ket{+}$ state projects the two optical modes into the desired Bell state,}
 &\rightarrow \frac{1}{2}\Big[(\ket{\alpha}_1 + \ket{-\alpha}_1)(\ket{\alpha}_2 + \ket{-\alpha}_2) + (\ket{i\alpha}_1 + \ket{-i\alpha}_1)(\ket{i\alpha}_2 + \ket{-i\alpha}_2)\Big].
 \end{align}

\end{widetext}

\null\clearpage

\bibliographystyle{mystyle}
\bibliography{references}


\end{document}